\documentclass[draft]{agujournal2019}
\usepackage{url} 
\usepackage{lineno}
\usepackage[finalnew]{trackchanges} 
\usepackage{soul}

\draftfalse

\journalname{JGR: Space Physics}

\usepackage{color}

\begin{document}

\title{Statistical study of energy dissipation in magnetic structures during turbulent reconnection in the Earth's magnetotail}

\authors{R. Wang\affil{1}, H. Ji\affil{1,2}, A. Robbins\affil{1,2}, K. Bergstedt\affil{1,2}, N. Ahmadi\affil{3}, R. Ergun\affil{3}, L.-J. Chen\affil{4}, J. Yoo\affil{2}, P. Shi\affil{2}, and Y. Doke\affil{5}}

\affiliation{1}{Princeton University, Princeton, NJ, USA}
\affiliation{2}{Princeton Plasma Physics Laboratory, Princeton, NJ, USA}
\affiliation{3}{Laboratory of Atmospheric and Space Sciences, University of Colorado Boulder, Boulder, CO, USA}
\affiliation{4}{NASA, Goddard Space Flight Center, Greenbelt, MD, USA}
\affiliation{5}{University of Tokyo, Tokyo, Japan}

\correspondingauthor{Hantao Ji}{hji@princeton.edu}
\correspondingauthor{Rachel Wang}{rachelwang@princeton.edu}

\begin{keypoints}
\item Magnetic field dissipation $\vec{j}\cdot\vec{E}$ in magnetic structures is dominated by electrons in the direction perpendicular to the magnetic field.
\item Energy exchange between fields and particles is statistically bidirectional with only a slight bias towards energy transfer from fields to particles.
\item 
In the electron energy equation under guiding center approximation, the Fermi term due to curvature drift in the electric field direction deposits the most energy.
\end{keypoints}

\begin{abstract}
Magnetic reconnection is a ubiquitous plasma phenomenon that plays a critical role in particle heating and energization. During reconnection, the topology of magnetic field rearranges, depositing energy into the surrounding plasma through bulk flow, thermal heating, or non-thermal particle acceleration. While the pathways of this transformation from magnetic energy into kinetic have been studied extensively in recent years through theoretical or case-by-case observations, comprehensive statistical studies remain limited. In this paper, we present a statistical investigation using data from the Magnetospheric Multiscale (MMS) mission, and detail the particle energization mechanisms in magnetic structures found near reconnecting regions in turbulent Earth's magnetotail. We find that electrons with motion perpendicular to the magnetic field dominate $\vec{j}\cdot\vec{E}$ dissipation. In contrast to the conventional picture of unidirectional energy transfer to particles by laminar two-dimensional (2D) reconnection, we find that energy exchange within magnetic structures during turbulent reconnection tends to be bidirectional with only a small positive bias from electromagnetic fields to particles. Specific electron energization mechanisms are quantified, including those due to parallel electric field, Fermi energization from curvature drift, betatron heating from magnetic field inhomogeneity, and polarization drift.
\end{abstract}

\section*{Plain Language Summary}
Magnetic reconnection is a fundamental process which converts magnetic energy into particle energy. The energy conversion process has been traditionally studied in detail on a case-by-case observational basis using 2D laminar reconnection models. In this paper, we use MMS observations to study this process statistically by using a large dataset of magnetic structures found during turbulent reconnection in the Earth's magnetotail. Our results show that, in contrast to the conventional picture of unidirectional energy transfer to particles by laminar 2D reconnection, energy exchange between the magnetic field and particles tends to be bidirectional. The same holds true for each of the proposed specific electron energization mechanisms (in decreasing order of significance): Fermi energization from curvature drift, parallel electric field, and betatron heating from magnetic field inhomogeneity.

\section{Introduction} \label{sec:intro}
\justifying

Magnetic reconnection is a ubiquitous and multi-scale physical process that converts magnetic field energy into kinetic energy of particles. Progress in understanding reconnection is being made from the perspectives of theory, simulation, laboratory experiment, and space observation \cite{yamada10,ji22}. Observations provide evidence of reconnection occurring in nature, prompting research into its roles in shaping and restructuring plasma dynamics of the environment. The Earth's magnetotail is one such natural laboratory to study the physics of reconnection through in-situ spacecraft measurements. Warped magnetic field lines meet on the nightside and form a neutral sheet parallel to the ecliptic plane extending several tens of Earth radii into the magnetotail. Two-dimensional laminar single X-line reconnection models are often employed, such as in \citeA{burch16} and \citeA{torbert18}, to interpret in-situ measurements of selected reconnection events involving magnetic structures. However, the magnetotail harbors magnetic structures of various configurations and parameters at all scales during turbulent reconnection, so a more statistical approach is needed to comprehensively characterize the multi-scale nature of reconnection. Recent studies have shown that turbulent reconnection is associated with significant heating and acceleration of particles up to several tens of keV per particle per second \cite{ergun20}.

NASA's Magnetospheric Multiscale (MMS) spacecraft is the most advanced spacecraft designed to study the Earth's magnetosphere; it has a temporal resolution of 8192 Sample/s for burst mode magnetic field measurements and 30 ms cadence for electron distributions, making it an ideal mission to study electron physics in turbulent reconnection. However, there have only been very few statistical studies of turbulent reconnection intervals using MMS data. \citeA{akhavan18} focused on the size distribution of Flux Transfer Events (FTEs) in the magnetosheath, following the first such attempt by \citeA{fermo11} using Cluster data. In the magnetotail, \citeA{bergstedt20} used an automated algorithm to detect different types of magnetic structures like plasmoids and current sheets, and studied their size distribution and magnetic energy dissipation. A more recent work has expanded the statistical study of size distribution to two different types of plasmoids and flux ropes \cite{smith24}. Magnetic energy dissipation in flux ropes has also been studied statistically in the turbulent magnetotail \cite{jiang21}, including various energization mechanisms \cite{ma22}. The latter has recently been extended to the magentosheath \cite{xu23}. In this paper, we present a statistical analysis of energy dissipation from over 700 magnetic structures measured aboard the MMS spacecraft in the Earth's magnetotail. We discuss their statistical properties in Section \ref{sec:statistics}, analyze the direction of energy transfer between field and particles in Section \ref{sec:dissipation}, and finally discuss the mechanisms of energy dissipation and its implications in Section \ref{sec:energy}.

\section{Data and Methodology} \label{sec:method}
\justifying

We used MMS magnetotail data measured in burst mode from June to August of 2017, including magnetic field measured by the Flux Gate Magnetometer (FGM) and electron moments measured by the Fast Plasma Investigator (FPI). Electron moments provided in the L2 data by the MMS team were occasionally unusable due to low electron density in some regions of the magnetotail. These regions exhibit unusually large spikes in electron velocity. We visually inspected all data used in this study to remove such intervals. The Geocentric Solar Ecliptic (GSE) coordinates are used for the rest of this paper.

Night side magnetotail intervals were visually selected from three months of data in 2017 based on the availability of burst mode data and occurrence of turbulent reconnection features indicated in previous studies \cite{ergun20}. The criterion for selection are turbulent electric and magnetic fields, flow reversal in ion bulk velocity, depletion of electron density, and heating and acceleration in ion and electron spectra.

The selection algorithm for magnetic structures (MSs) is adapted from \citeA{bergstedt20}. The algorithm identifies zero crossings of $B_z$ as locations of potential MSs. The Maximum Variance Analysis (MVA) is applied to a short interval surrounding each crossing to rotate it into the local coordinate system of the interval. If the zero crossing is still present after MVA rotation, and the magnetic field is bipolar, then it is selected as an MS. The MVA analysis is repeated on MMS2-4 to ensure the same structure is observed on all four spacecraft, then multi-spacecraft timing method is used to compute the velocity and spatial scale of each MS. Based on the polarity of ion bulk velocity, current density, and direction of bipolar signature in $B_z$, MSs are categorized into plasmoids with hoop-like magnetic configuration, push current sheets between two plasmoids pushing into each other, pull current sheets between plasmoids which are not actively merging, or undetermined \cite{bergstedt20}. The selection procedure is automated without the need for visual inspection, but confirmed visually afterwards to ensure consistency and impartiality.

\begin{figure}
\includegraphics[width=1.0\linewidth,angle=0]{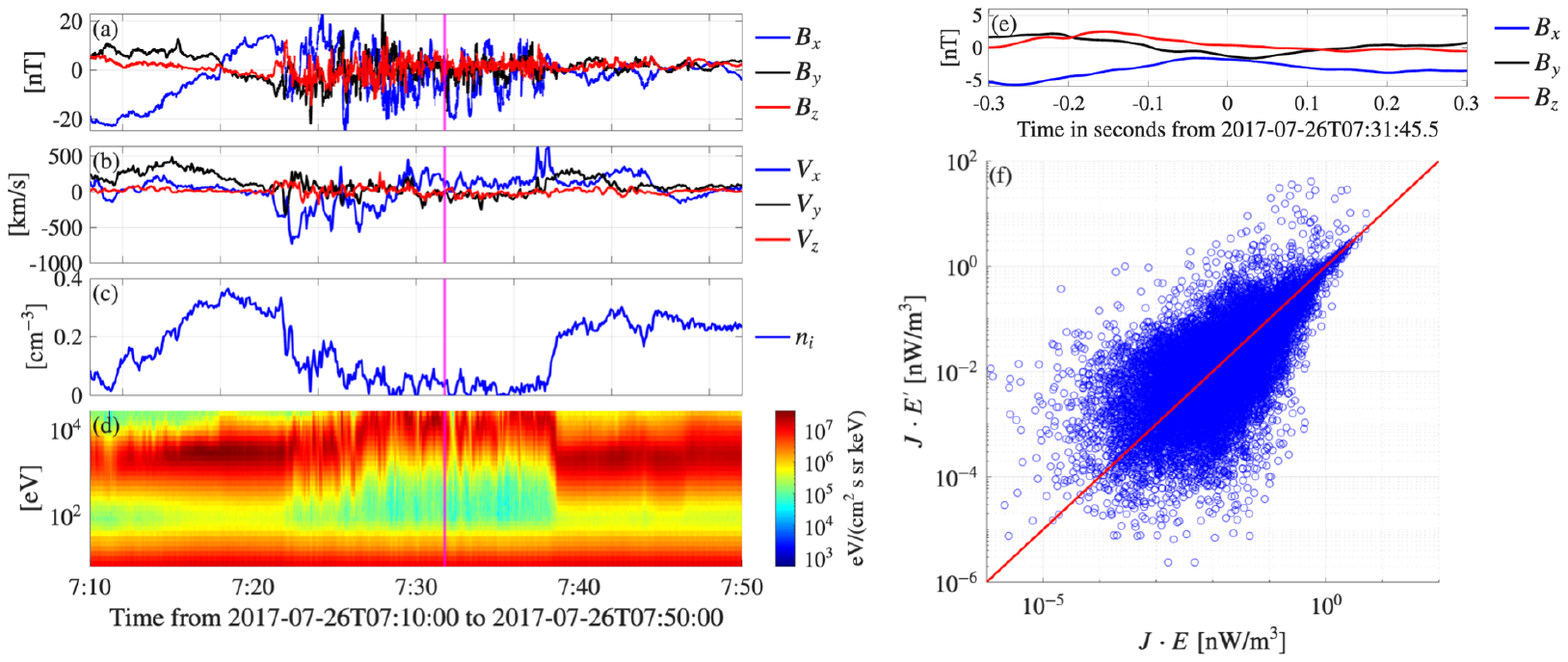}
\caption{Overview of an interval of magnetotail reconnection on 26 July 2017: (a) magnetic field measured aboard MMS1 in GSE coordinates; (b) ion bulk velocity in GSE showing flow reversal; (c) electron density; (d) omnidirectional electron spectrum showing electron heating and acceleration; (e) magnetic field of an MS selected in this interval; (f) comparison between $\vec{J}\cdot\vec{E}$ and $\vec{J}\cdot\vec{E'}$ with $y=x$ line indicated in red. Magenta lines in all panels indicate center of the MS from panel (e), identified by the algorithm, which is also featured in Figure \ref{fig8} (a)-(c).}
\label{fig1}
\end{figure}

Figure \ref{fig1} presents an overview of a selected interval in this study using measurements from MMS1. Measurements aboard other MMS spacecraft look qualitatively similar. Panels (a)-(c) show enhanced turbulence in the magnetic field, reversal in ion bulk flow, and depletion of electron density respectively. All of these features are indicative of active magnetotail reconnection \cite{torbert18}. The electron omnidirectional spectrum in panel (d) shows signatures of both electron heating and acceleration. In this interval, the MMS spacecraft were most likely traversing through a reconnecting tail current sheet as demonstrated in \citeA{ergun20}. Magenta lines in this figure indicate the location of the MS shown in panel (e), this is the same MS as Figure \ref{fig8} (a)-(c). Finally, panel (f) compares $\vec{J}\cdot\vec{E}$ with $\vec{J}\cdot\vec{E'}$ for this interval, where the first quantity is dissipation in the spacecraft frame, and the second is in the electron frame $\vec{E'}=\vec{E}+\vec{v_e}\times\vec{B}$. Here, current density $\vec{J}$ is calculated from the Level 2 moment. These two quantities have good agreement, consistent with the assumption that the majority of MSs are electron-scale structures \cite{chen19}. The similarity of dissipation in spacecraft and electron frames imply that energy transfer to particles is entirely due to non-ideal electric fields, and that energy conversion is frame-independent. Since $\vec{J}\cdot\vec{E}$ and $\vec{J}\cdot\vec{E'}$ are very similar, we will use $\vec{J}\cdot\vec{E}$ for the rest of this paper.

\section{Statistical Overview} \label{sec:statistics}

\begin{figure}
\includegraphics[width=1.0\linewidth,angle=0]{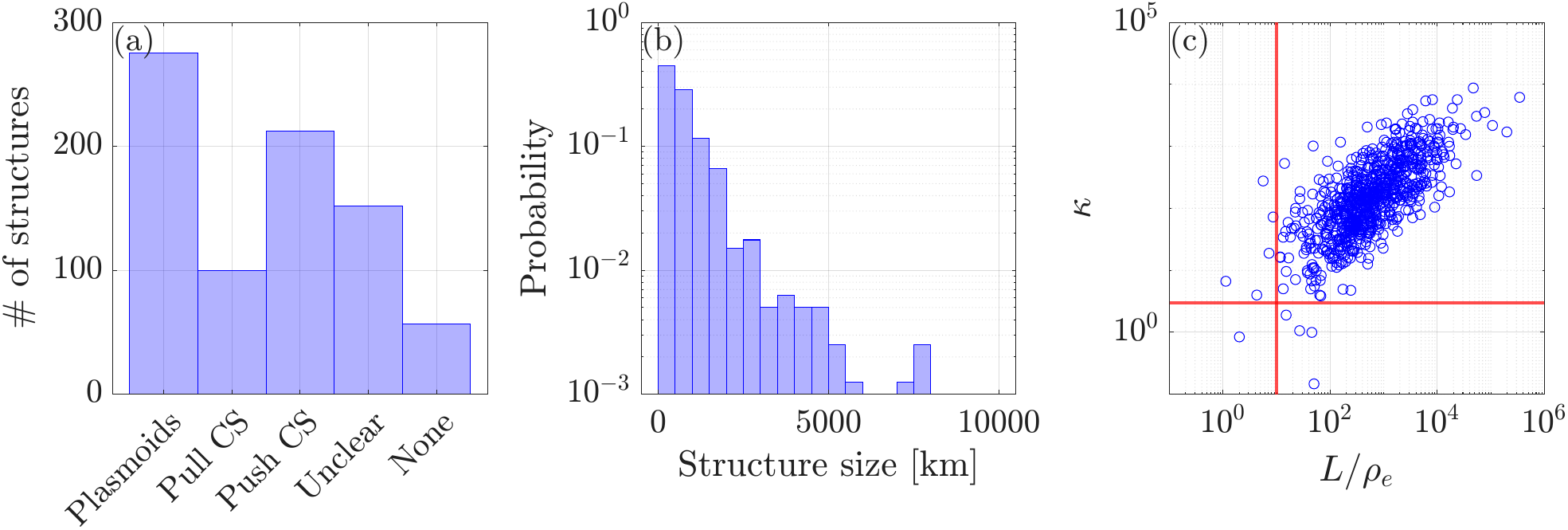}
\caption{Statistical summary of MS properties: (a) distribution showing number of MSs classified by type; (b) log-linear probability distribution of spatial scale of MSs; (c) events selected based on $\kappa=\sqrt{R_c/R_L}$ and spatial scale normalized by electron gyroradius.}
\label{fig2}
\end{figure}

Figure \ref{fig2} gives an overview of some statistical properties of MSs. Panel (a) shows the classification of MSs using the algorithm from \citeA{bergstedt20}. MSs are classified as plasmoid-like or current sheet-like; current sheet-like structures are further divided into push or pull types depending on whether they are formed from two converging or diverging plasmoids. The algorithm selected 796 MSs, where 35\% are plasmoids, 12\% are pull current sheets, 27\% are push current sheets. The rest are undetermined, which fit either into multiple categories (labelled ``Unclear") or none (labelled ``None").

Panel (b) shows the spatial scale of MSs, approximated as the product of velocity and temporal duration of MSs, where the velocity of structures in the normal direction determined by MVA was computed using the Spatio-Temporal Difference (STD) method \cite{shi06}. The sizes of MSs range from hundreds to thousands of kilometers, so most MSs are electron scale rather than ion scale. The spatial scales of MSs roughly follow a decaying exponential consistent with previous studies \cite{fermo11,akhavan18,bergstedt20} and theoretical prediction \cite{fermo10}. However, from this distribution alone, it is difficult to confidently distinguish between an exponential and a power law, because their difference is the most obvious at the tail of the distribution consisting of large scale structures, which are often overlooked by the selection algorithm.

Panel (c) shows both the $\kappa$ parameter and structure sizes normalized by electron gyroradius. The selected events are in the upper right quadrant with boundaries marked in red. $\kappa$, also called the adiabaticity parameter~\cite{buechner89} is the square root of the ratio between curvature radius and Larmor radius, calculated by finding the spatial derivative of the magnetic field with four-spacecraft methods. All events with $\kappa < 3$ were discarded from the dataset to remove error from non-adiabatic motion of electrons and to allow for the application of guiding center approximation~\cite{steinvall2025}. Such approach has previously been used in other studies such as~\citeA{ma22}. Since most structures fall above the red $\kappa=3$ line, it is justified to assume that electrons are magnetized but ions are unmagnetized in this analysis. Less than 5\% of the events fall below either red line and are discarded. Note that due to the limitations of the selection algorithm, larger scale structures are more likely to be excluded from having ambiguous boundaries due to random fluctuations, so the algorithm preferentially selects electron scale structures, therefore this study will also give preference to analyzing electron behaviors.

\section{Partition of Energy Dissipation} \label{sec:dissipation}

\begin{figure}
\includegraphics[width=1.0\linewidth,angle=0]{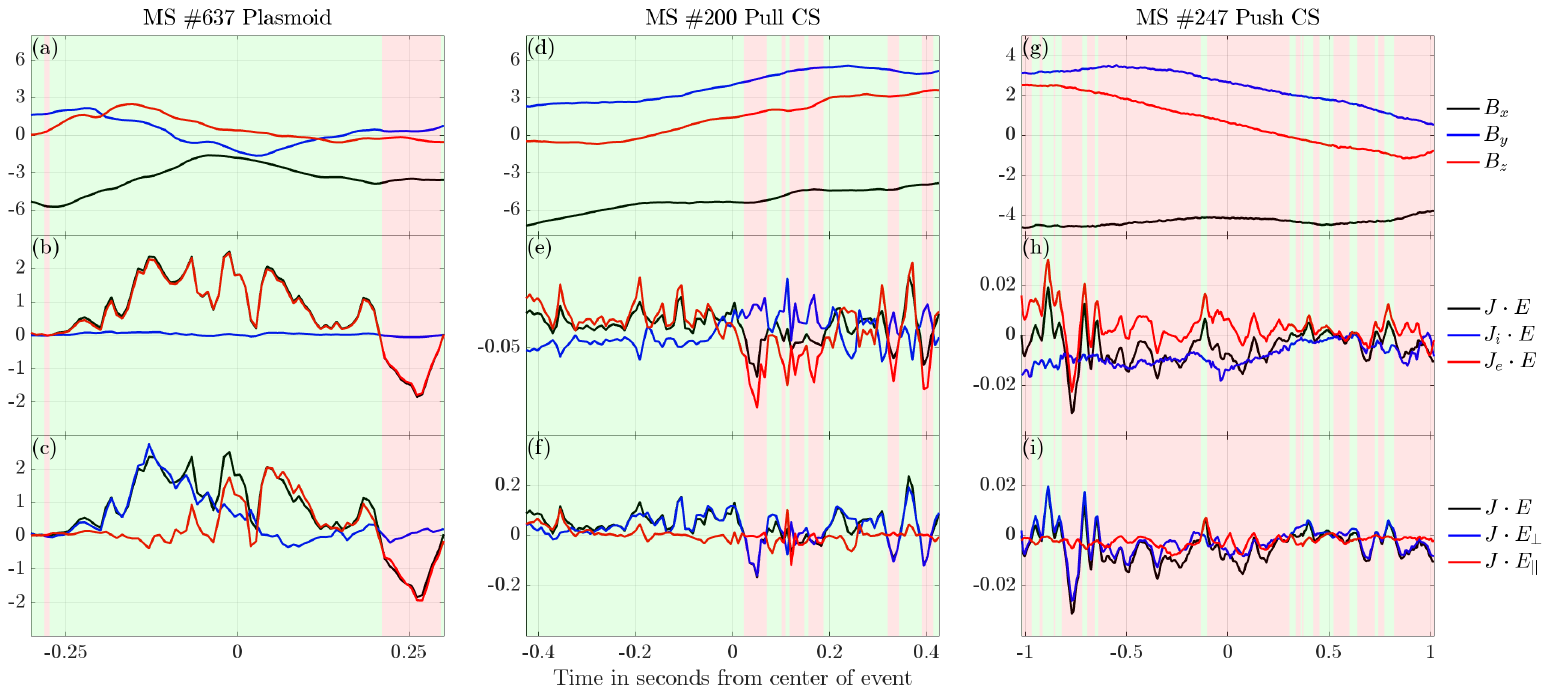}
\caption{Three MS examples with different dissipation patterns, the plot background is green if total $J\cdot E$ is positive, and red if negative. (a) magnetic field in GSE coordinates; (b) ion, electron, and total dissipation; (c) perpendicular, parallel, and total disspation; (d)-(f) same as (a)-(c) for a pull CS; (g)-(i) same as (a)-(c) for a push current sheet.}
\label{fig8}
\end{figure}

There is strong evidence in Figure \ref{fig1} of electron heating and acceleration inside an MS. To understand the process of energy dissipation into particles, we compare the various contributing components of $\vec{J}\cdot\vec{E}$ with a method inspired by \citeA{wilder18} to quantify whether perpendicular, parallel, electron, or ion dissipation dominates. Figure \ref{fig8} shows three examples of MSs that are representative of dissipation patterns inside structures. While conventional reconnection events typically involve sustained positive dissipation, the same is not true for these MSs, most of which are not actively reconnecting. Energy exchange is thus mediated not by reconnection but by turbulent fluctuations which both convert magnetic energy into particle energy and vice-versa. To offer a concrete picture of the bidirectional nature of energy dissipation, we present three case studies comprising one MS of each type.  Consider a typical MS, shown in Figure \ref{fig8}(g)-(i), a push current sheet. The energy dissipation is clearly net negative. However, towards the beginning and end of the interval, there are alternating periods of positive dissipation. Furthermore, note that the ions and electrons do not necessarily gain and or lose energy at the same time. At $t = -$0.75 s, both species lose energy to the field while for most of the next second the electrons gain energy while the ions lose energy. This transfer of energy from one species to another has also been observed in laboratory experiments \cite{bose2024}. Potentially, this can also be due to structure motion, which is quantified in the next figure.  Note also that there is no obvious correlation between the fluctuations and their positions within the magnetic structure.

The plasmoid MS in Figure \ref{fig8}(a)-(c) shows a converse case where the dissipation is mostly positive but transitions into negative by the end. In this case, the electron term dominates. The pull current sheet MS in Figure \ref{fig8}(d)-(f) starts out nearly entirely with positive dissipation until just after $t = $ 0 s, whereupon the energy begins sloshing back and forth between the fields and the particles. These cases are typical; nearly all events involve some periods of positive and negative dissipation as energy is transferred between electrons, ions, and fields. This randomness extends to the perpendicular and parallel components where the same disorder is visible. In summary, while these MSs are clearly distinguished from the surrounding plasma by higher intensity of dissipation, this dissipation needs not be orderly nor coherent. Thus, a statistical approach to analyze the direction and particles that dominate dissipation is needed.

\begin{figure}
\includegraphics[width=1.0\linewidth,angle=0]{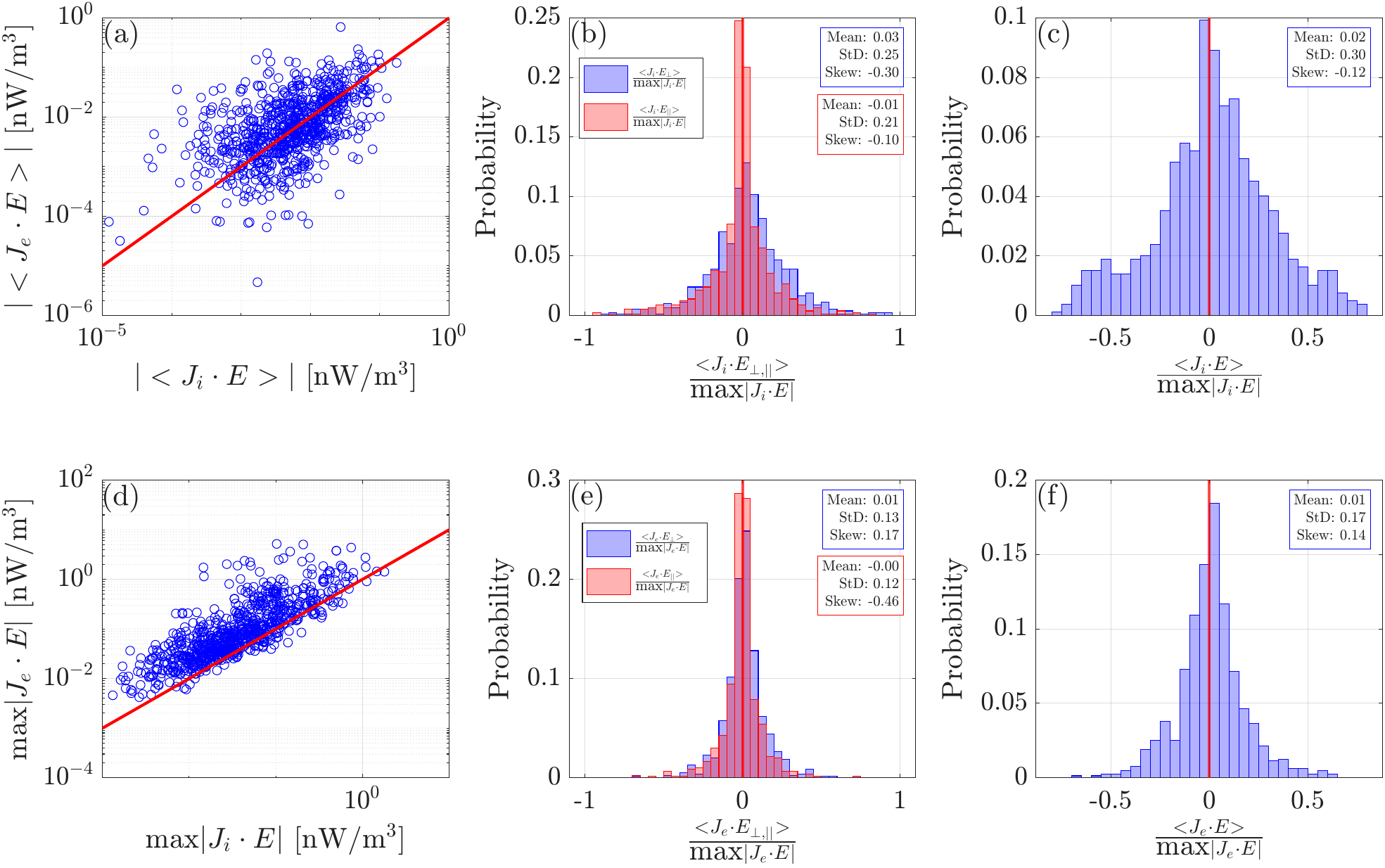}
\caption{(a) comparison between averaged $\vec{j_i}\cdot\vec{E}$ and $\vec{J_e}\cdot\vec{E}$ with $y=x$ line indicated in red; (d) comparison between maximum magnitude of $\vec{J_i}\cdot\vec{E}$ and $\vec{J_e}\cdot\vec{E}$ with $y=x$ line indicated in red; (b),(e) probability distribution of average perpendicular and parallel $\vec{j_i}\cdot\vec{E}$ and $\vec{j_e}\cdot\vec{E}$ normalized by their respective maximum magnitude; (c),(f) probability distribution of average total $\vec{j_i}\cdot\vec{E}$ and $\vec{j_e}\cdot\vec{E}$ normalized by total $\vec{j}\cdot\vec{E}$. For each distribution, the mean, the standard deviation, and the skewness are calculated and also shown.}
\label{fig3}
\end{figure}

We start by showcasing the differences between two methods of comparison. Figure \ref{fig3}(a) and (d) compare electron vs ion contributions to $\vec{J}\cdot\vec{E}$, but in (a) the quantity in question is averaged over each MS, whereas in (d) and the maximum of the absolute value is used. While both methods are valid and justified comparisons of the scale of dissipation, they give distinctly different results. (a) shows that dissipation in both species are similar in magnitudes when averaged, and (d) shows that electron dissipation has much larger peaks. The structure motion contributes to a common velocity for the particles, hence giving the agreement in (a) between $\vec{J_i}\cdot\vec{E}$ and $\vec{J_e}\cdot\vec{E}$, however, (d) shows that inside these structures, electron energy dissipation fluctuates much more strongly than ions.

Panels (b) and (e) show comparisons between perpendicular vs parallel dissipation for ions and electrons, respectively. Perpendicular dissipation is in blue, and parallel is in red. The choice to normalize by $\max{|\vec{J}\cdot\vec{E}|}$ scales the distribution to always range from -1 to 1 for ease of comparisons. Normalizing by $\langle\vec{J}\cdot\vec{E}\rangle$ is not ideal, because in some MSs the total contribution from $\vec{J}\cdot\vec{E}$ is close to zero, despite having non-zero directional contributions.

The broader distributions for perpendicular dissipation in both cases, compared to parallel dissipation which is more strongly peaked at zero, show that a larger amount of energy is usually dissipated into perpendicular particles for both ions and electrons. The mean of every distribution is an order of magnitude smaller than the standard deviation, so all four distributions are qualitatively symmetric about zero, demonstrating that statistically net energy exchange is balanced between fields and particles. Even though in some events particles could be strongly accelerated or heated, these extreme cases are rare and most MSs have minimal net dissipation. Energy exchange is more bidirectional than unidirectional between fields and particles. 

In 2D laminar reconnection, parallel electric fields are usually the primary means of dissipation~\cite{Zenitani2011}. However, in turbulent 3D reconnection, perpendicular disspation often dominates \cite{Pyakurel2025}. Stronger dissipation in the perpendicular direction than parallel implies that particles are energized primarily by perpendicular motion of electrons, such as curvature-driven Fermi acceleration and betatron heating from conservation of magnetic moment, rather than by direct acceleration by parallel electric fields. Note that Fermi acceleration mechanism energizes particle’s parallel energy while betatron heating mechanism energizes particle’s perpendicular energy. This verifies our claim that the 2D laminar model commonly used is not suitable for turbulent or 3D reconnection environments such as the magnetotail, since energy conversion is caused by complex field geometries rather than simple 2D structures.

Panels (c) and (f) combine both distributions in the previous panels and further demonstrate this idea of bidirectional energy exchange. They show the distributions of average total dissipation $\vec{J}\cdot\vec{E}$ for ions and electrons, normalized by their respective maximum inside the structure. The symmetry of this quantity about zero indicates that net energy transfer between fields and particles is statistically very small. Despite the presence of events where a large amount of energy is exchanged, such as those in the tail ends of the distributions, the majority of MSs have minimal net energy transfer, so energy goes back-and-forth between fields and particles.

\begin{figure}
\includegraphics[width=1.0\linewidth,angle=0]{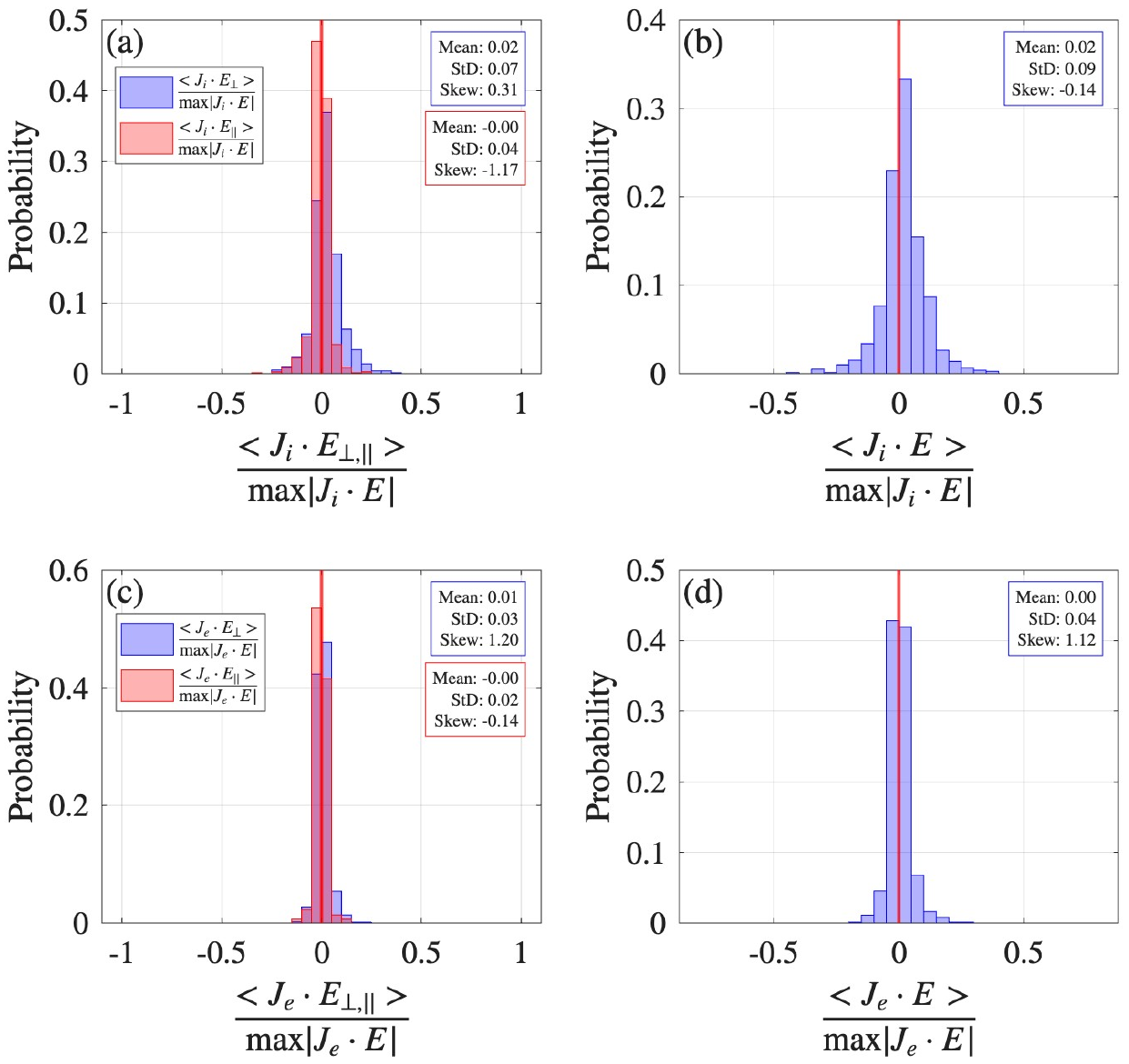}
\caption{Distributions similar to Figure \ref{fig3} of dissipation outside MSs: (a),(c) probability distribution of average perpendicular and parallel $\vec{j_i}\cdot\vec{E}$ and $\vec{j_e}\cdot\vec{E}$ normalized by their respective maximum magnitude; (b),(d) probability distribution of average total $\vec{j_i}\cdot\vec{E}$ and $\vec{j_e}\cdot\vec{E}$ normalized by total $\vec{j}\cdot\vec{E}$. For each distribution, the mean, the standard deviation, and the skewness are calculated and also shown.}
\label{fig7}
\end{figure}

The conventional picture of energy dissipation in reconnection is from fields to particles, so the net energy exchange being close to zero prompts the question of whether the selected MSs are physically-significant coherent structures or merely random turbulence. Figure \ref{fig7} reproduces panels (b), (c), (e), and (f) from Figure \ref{fig3} using intervals surrounding every MS with ten times the duration, not including the MS itself. This allows a direct comparison of dissipation behavior inside versus outside of MSs. The most prominent feature in Figure \ref{fig7} for dissipation outside MSs is that every distribution strongly peaks at zero, indicated again by the fact that the mean is an order of magnitude smaller than the standard deviation. This implies that the average dissipation outside of an MS is generally insignificant, consistent with the assumption that it is mostly random turbulence. On the other hand, the same distributions in Figure \ref{fig3} within MSs are more spread out and have longer tails. Therefore, the majority of MSs dissipate energy consistently within its boundary, and the magnitude is physically significant compared to random turbulence. This comparison between inside and outside of MSs confirms the accuracy of the selection algorithm, and proves that zero net energy exchange happens not only in random turbulence, but also inside coherent structures in reconnection.

\begin{figure}
\includegraphics[width=1.0\linewidth,angle=0]{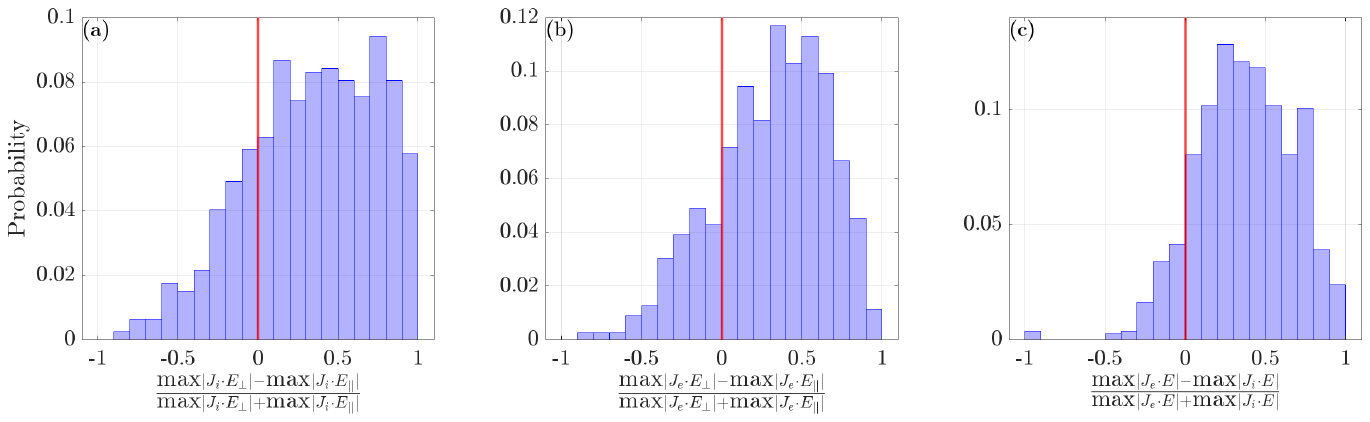}
\caption{(a)-(b) distributions of perpendicular vs. parallel dissipation ratio for electrons and ions; (c) distribution of electron vs. ion dissipation ratio.}
\label{fig4}
\end{figure}

Figure \ref{fig4} (a) and (b) quantify conclusions drawn from Figure \ref{fig3} with distributions of the perpendicular to parallel dissipation ratio, defined as:
$$\frac{\max{\left|\vec{J}_{\perp}\cdot\vec{E}_{\perp}\right|}-\max{\left|J_{||}E_{||}\right|}}{\max{\left|\vec{J}_{\perp}\cdot\vec{E}_{\perp}\right|}+\max{\left|J_{||}E_{||}\right|}}$$
which is a modified version of a similar ratio used in \citeA{wilder18}. This modification allows the dissipation ratio to be normalized to a value between -1 and 1. In a structure where perpendicular dissipation completely dominates, this quantity simplifies to 1 because the parallel term is negligible. Similarly, the ratio becomes -1 if dissipation is completely parallel. When perpendicular and parallel dissipation are comparable in magnitudes, this fraction becomes zero. The absolute value allows perpendicular and parallel quantities to be compared based on magnitudes alone, regardless of energy flow into or out of particles.

The histogram in both panels are skewed to the right, indicating that perpendicular dissipation dominates in most structures in both ions and electrons. This agrees with the message from Figure \ref{fig3}(b) and (e), that the largest energy exchange between fields and particles occurs with perpendicular particle motions.

In Figure \ref{fig4}(c) we compare contributions between dissipation to electrons and ions. We define the electron vs ion dissipation ratio in an analogous way to Figure \ref{fig4}(a) and (b), 
$$\frac{\max{\left|\vec{J}_e\cdot\vec{E}\right|}-\max{\left|\vec{J}_i\cdot\vec{E}\right|}}{\max{\left|\vec{J}_e\cdot\vec{E}\right|}+\max{\left|\vec{J}_i\cdot\vec{E}\right|}}.$$
Electron-dominated dissipation results in a ratio of 1, ion-dominated gives -1, and the ratio becomes 0 if they are comparable in magnitudes. The distribution in Figure \ref{fig4}(c) is strongly skewed towards the right, which means that electrons dominate the energy exchange. This is in agreement with Figure \ref{fig3}(a) (d): while both species on average dissipate similar amounts of energy, electrons have much larger spikes of energy exchange with the field compared to ions.

To summarize the implications of the previous figure: perpendicular dominated dissipation for both electrons and ions indicates a departure from the classical 2D laminar reconnection model, where parallel electric fields play the most important role in electron dissipation, and ions are energized by bulk flow. These MSs verify the occurrence of 3D turbulent reconnection happening in the magnenotail and the need of a more complex reconnection model.

\section{Electron energy deposition mechanisms} \label{sec:energy}

While $\vec{J}\cdot\vec{E}$ gives the total dissipation inside MSs, it does not tell us the mechanisms or pathways of energy transfer. Energization rate of electrons can be decomposed into components using the guiding center approximation \cite{northrop63,dahlin14}. The same approach is used in \citeA{sun22}. This approximation is justified because the spatial scales of MSs are generally much larger than electron gyroradius, so electrons are magnetized, and all events have $\kappa \geq 3$, see Section \ref{sec:method}. The equation is reproduced below:
$$\frac{dU}{dt}=E_{||}J_{||}+\frac{p_{\perp}}{B}\frac{dB}{dt}+\left(p_{||}+m_en_ev_{||}^2\right)\vec{v}_E\cdot\left(\vec{b}\cdot\nabla\vec{b}\right)$$
where $U$ is the kinetic energy of electrons, and the terms on the right hand side represent energization due to parallel electric field, Betatron heating due to conservation of magnetic moment, Fermi mechanism due to field-line curvature drift. $\vec{v}_E$ is the $\vec{E}\times\vec{B}$ drift velocity and $d/dt\equiv \partial/\partial t+ \vec{v}_E\cdot \nabla$. In terms of their relation with field dissipation discussed in Section \ref{sec:dissipation}, the first term is identical to $J_{||}E_{||}$, and the other two are parts of $J_{\perp}E_{\perp}$.

Fermi acceleration deposits energy into electrons via curvature drift. It acts through the parallel motion of electrons~\cite{drake06}. Betatron heats particles by gradient of magnetic field strength and contributes to perpendicular energy \cite{Montag2017}. Polarization drift $\frac{1}{2}m_en_e\frac{d}{dt}|\vec{v}_E|^2$ also contributes to electron energization~\cite{Li2015,Li2017} and occurs due to the presence of a time varying electric field. Because of the scaling of polarization current with particle mass, this term is not as important for electrons. It could become more significant in fields that have large temporal variation, such as when the turbulent plasma experiences rapid fluctuations, or in reconnection exhausts \cite{Eriksson2024E}. In these MSs in turbulent reconnection region in the magnetotail, field geometry is more important in terms of dissipation. This is common in large scale magnetotail reconnection or solar wind reconnection. Given the typical parameters inside MSs in the Earth's magnetotail, the polarization drift term is two orders of magnitudes smaller than the other two. Therefore, it is not included in the electron energization calculations.

\begin{figure}
\includegraphics[width=1.0\linewidth,angle=0]{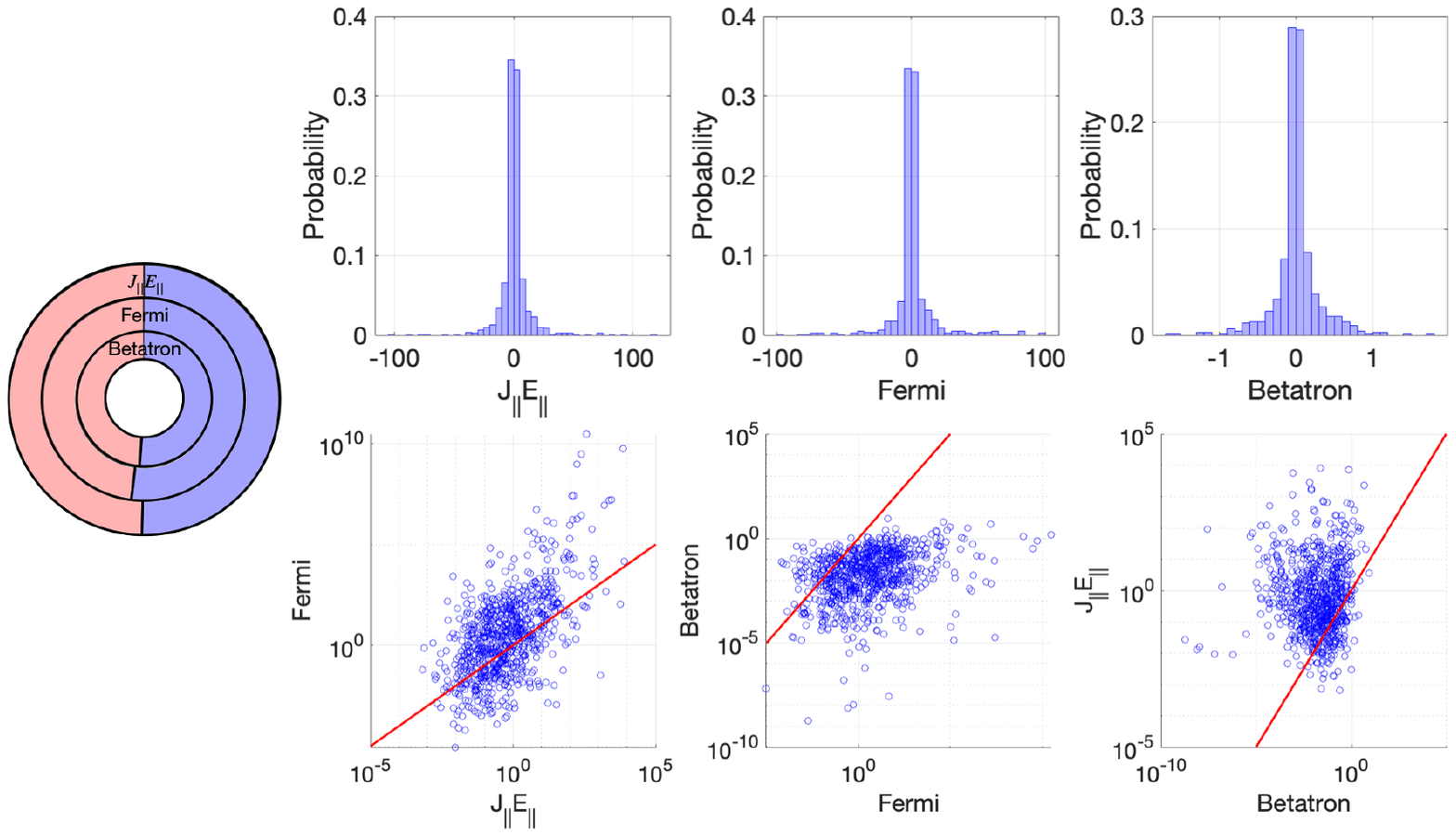}
\caption{(a) stacked pie charts showing sign distribution of $j_{||}E_{||}$, Fermi, and betatron terms, where blue is positive and red is negative; (b)-(d) probability distributions of each term, (e)-(g) relative size comparisons between different terms, with red line indicating $y=x$ in every panel.}
\label{fig5}
\end{figure}

Figure \ref{fig5}(a) demonstrates the sign distribution of each term: $J_{||}E_{||}$, Fermi, and betatron, respectively. We assume that the temporal derivative $\partial /\partial t$ is negligible compared to the convective derivative in accordance with common practice. Every term is integrated over the duration of the MS to determine the sign of the net contribution. The blue section is positive, where energy flows into the particles from the fields, and the red section is negative. 

$J_{||}E_{||}$ does not display a statistically significant bias in the sign. This shows that energy is equally likely to be given to and taken from the electrons by parallel electric field and the net energy transfer is small, despite some MSs having large $J_{||}E_{||}$. Statistically, the energy moves bidirectionally between particles and field, and unidirectional cases of energy transfer are rare. Both Fermi and betatron terms are slightly biased positive, indicating that both of these mechanisms are responsible for net energy exchange into the particles statistically. 

The distributions in Figure \ref{fig5}(b)-(d) show sign and size of variance of the energy terms. $J_{||}E_{||}$ and Fermi terms are comparable in magnitudes, whereas betatron heating is much smaller statistically. Figure \ref{fig5}(e)-(g) showcase size comparisons between terms explicitly. Due to log scales used for these plots, all terms are shown in absolute value and the sign information is lost.

Summarizing the contributions from each term: Fermi term deposits the most amount of energy into electrons, due to its large magnitude and bias towards the positive side. While $J_{||}E_{||}$ is equally large in magnitude, there is very little net energy transfer since the distribution is more symmetric about zero. Finally, even though betatron term has a positive bias, it is small in magnitude because the effect is dependent on pitch angle. The total energy transfer into electrons due to betatron heating is positive but small. If reliable data were available at higher energy, the positive bias could be more apparent \cite{dahlin17}.

The dominance of the Fermi term implies that curved magnetic field lines and contraction of field structures is more important than field line compression in these MSs. The system is non laminar and exhibit lots of dynamically evolving structures such as flux ropes and MSs. This is typical of turbulent reconnection regions in the magnetotail such as the ones studied in this paper. Similar conclusions have also been reached through simulations \cite{dahlin14} and MMS observations \cite{Li2015}. This has implications such as more efficient energy transfer, due to Fermi processes being able to energize particles more rapidly, compared to betatron which is more localized heating. 

The consequence of each energization mechanism should theoretically correlate with changes in electron temperature or power law index of the high energy tail. However, due to the large uncertainty in the power law tail, and significant influence of stochastic heating on electron temperature, we were unable to justify the correlation between temperature difference and size of energy terms as purely causal.

\section{Summary and Discussion}

Most research on magnetic structures near reconnecting regions are case studies of 2D laminar reconnection, which represents extremely energetic events of dissipation from magnetic fields to particles. While they reveal the physics of conventionally studied magnetic reconnection, the amount of net energy dissipated in these events is usually large but not representative of MSs, see Section \ref{sec:dissipation}. Statistically, the majority of MSs dissipate little energy into particles after integration. The net energy exchange is close to zero regardless of particles or direction of dissipation. This is due to the fluctuating nature of $\vec{J}\cdot\vec{E}$. Energy is exchanged in both directions between field and particles in MSs, as opposed to the conventional picture of unidirectional energy flow during 2D laminar reconnection from field to particles. 

In laminar reconnection, $\vec{J}\cdot\vec{E}$ is strictly positive, and dissipation is irreversible. Because of this, laminar model should be very carefully applied to study turbulent reconnection in the magnetotail, where energy transfer is intermittent and energy is deposited both into the plasma and the particles. Symmetric $\vec{J}\cdot\vec{E}$ distributions have previously been linked to reversible energy transfer in simulations of kinetic turbulence transfer\cite{Wan2015}, verified both in theory and numerically \cite{Cerri2021}, and measured by the MMS in the magnetosheath \cite{Bandyopadhyay2020}.
Occurrences of structures with large dissipation of field energy is not uncommon, and these more extreme events on the tail ends of the $\vec{J}\cdot\vec{E}$ distributions are what we commonly see in case studies. Therefore, these results do not contradict with previous studies, but rather complement them. 

We showed that Fermi term has the largest variance inside MSs, meaning that it usually manifests as large spikes in energy. The implications regarding this could be rather profound \cite{lemoine22}, albeit we will restrict the discussion to the context of reconnection. The Fermi mechanism is generally known to be responsible for non-thermal particle acceleration, which we could not observe directly due to large uncertainties in the slope of the power law tail that are often comparable to or exceeding the effect of Fermi acceleration. Nonetheless, some MSs exhibit more extreme energy dissipation, thus there is a large enough change in high energy electrons and the power law index which exceeds the uncertainty, making case studies possible. Such cases are not the focus of this study.

A breakdown of specific energization mechanisms shows that Fermi acceleration driven by particle reflection from curved field lines dominates over both betatron heating and acceleration by parallel electric fields. This implies that dynamics of magnetic field line is more important than compression or temporal variation in energy conversion. Fermi acceleration has been shown to be very efficient in reconnection exhausts and turbulent flux ropes, where particles follow curved field lines and become energized by curvature drift \cite{dahlin14, Li2015}. On the other hand, betatron heating associated with strengthening magnetic field is only secondary, showing relatively weak magnetic compression in these MSs. We also find negligible energy contribution from polarization drift, which demonstartes the lack of rapid temporal variation in electric fields. While polarization effects are not important here compared to other terms, they become more relevant where high-frequency fluctuations are prevalent.

To summarize, these results suggest that energy dissipation driven by turbulent reconnection in the magnetotail is determined by a combination of field geometry and compression, less so on temporal variation, with Fermi acceleration being the dominant process. The relative weakness of betatron and parallel electric field contributions emphasizes the importance of 3D turbulent field structures in electron energization, and the insufficiencies of modeling magnetotail reconnection with 2D laminar models.

While this work focused on magnetic structures collectively, there are of course important differences between plasmoids, pull current sheets and push current sheets. Preliminary analysis of our data suggests that plasmoids are statistically significantly likely to dissipate magnetic energy on net while pull current sheets conversely undergo negative dissipation. This could offer clues as to the sensitivity of turbulent fluctuations to the local current sheet geometry. We leave to further research the task of assessing how the results presented herein depend on the particular type of magnetic structure under consideration.

Statistical investigations such as this study open up potential for future works to explore a turbulent 3D reconnection model, which are more relevant to naturally occurring plasmas than the commonly used 2D laminar picture. It would also be of interest to compare these results to electron heating in other environments such as the Earth's magnetosheath~\cite{xu23}. Recent developments in automated algorithms \cite{bergstedt20} and machine learning models \cite{bergstedt24} to identify structures of interests would also greatly simplify the work involved in obtaining a robust and extensive dataset for statistical analysis. An unexplored question that could be answered by future statistical works is the conditions under which the 2D laminar picture transitions into turbulent 3D reconnection. Answering this not only helps us understand the kinetic physics inside the magnetotail, but also gives valuable insight into the inner workings of astrophysical reconnection in general.

\section*{Data Availability Statement}
The data used in this research is publicly available at \url{https://lasp.colorado.edu/mms/sdc/public/}.

\section*{Conflict of Interest Statement}
The authors have no conflicts of interest to disclose. 

\acknowledgments
We thank Xiaocan Li for helpful comments on the manuscript. This work was supported by NASA under Grants No. NNH15AB29I and 80HQTR21T0105, and by the U.S. Department of Energy's Office of Fusion Energy Sciences under Contract No. DE-AC0209CH11466. We thank the MMS and PySPEDAS teams for excellent data and analysis package.

\newpage

\begin{thebibliography}{}

\bibitem [\protect \citeauthoryear {%
Akhavan-Tafti%
\ \protect \BOthers {.}}{%
Akhavan-Tafti%
\ \protect \BOthers {.}}{%
{\protect \APACyear {2018}}%
}]{%
akhavan18}
\APACinsertmetastar {%
akhavan18}%
\begin{APACrefauthors}%
Akhavan-Tafti, M.%
, Slavin, J\BPBI A.%
, Le, G.%
, Eastwood, J\BPBI P.%
, Strangeway, R\BPBI J.%
, Russell, C\BPBI T.%
\BDBL {}Burch, J\BPBI L.%
\end{APACrefauthors}%
\unskip\
\newblock
\APACrefYearMonthDay{2018}{}{}.
\newblock
{\BBOQ}\APACrefatitle {MMS Examination of FTEs at the Earth's Subsolar
  Magnetopause} {Mms examination of ftes at the earth's subsolar
  magnetopause}.{\BBCQ}
\newblock
\APACjournalVolNumPages{Journal of Geophysical Research: Space
  Physics}{123}{2}{1224-1241}.
\newblock
\begin{APACrefDOI} \doi{10.1002/2017JA024681} \end{APACrefDOI}
\PrintBackRefs{\CurrentBib}

\bibitem [\protect \citeauthoryear {%
{Bandyopadhyay}%
\ \protect \BOthers {.}}{%
{Bandyopadhyay}%
\ \protect \BOthers {.}}{%
{\protect \APACyear {2020}}%
}]{%
Bandyopadhyay2020}
\APACinsertmetastar {%
Bandyopadhyay2020}%
\begin{APACrefauthors}%
{Bandyopadhyay}, R.%
, {Matthaeus}, W\BPBI H.%
, {Parashar}, T\BPBI N.%
, {Yang}, Y.%
, {Chasapis}, A.%
, {Giles}, B\BPBI L.%
\BDBL {}{Burch}, J\BPBI L.%
\end{APACrefauthors}%
\unskip\
\newblock
\APACrefYearMonthDay{2020}{{\APACmonth{06}}}{}.
\newblock
{\BBOQ}\APACrefatitle {{Statistics of Kinetic Dissipation in the Earth's
  Magnetosheath: MMS Observations}} {{Statistics of Kinetic Dissipation in the
  Earth's Magnetosheath: MMS Observations}}.{\BBCQ}
\newblock
\APACjournalVolNumPages{Phys. Rev. Lett.}{124}{25}{255101}.
\newblock
\begin{APACrefDOI} \doi{10.1103/PhysRevLett.124.255101} \end{APACrefDOI}
\PrintBackRefs{\CurrentBib}

\bibitem [\protect \citeauthoryear {%
{Bergstedt}%
\ \BBA {} {Ji}%
}{%
{Bergstedt}%
\ \BBA {} {Ji}%
}{%
{\protect \APACyear {2024}}%
}]{%
bergstedt24}
\APACinsertmetastar {%
bergstedt24}%
\begin{APACrefauthors}%
{Bergstedt}, K.%
\BCBT {}\ \BBA {} {Ji}, H.%
\end{APACrefauthors}%
\unskip\
\newblock
\APACrefYearMonthDay{2024}{}{}.
\newblock
{\BBOQ}\APACrefatitle {{A Novel Method to Train Classification Models for
  Structure Detection in In-situ Spacecraft Data}} {{A Novel Method to Train
  Classification Models for Structure Detection in In-situ Spacecraft
  Data}}.{\BBCQ}
\newblock
\APACjournalVolNumPages{Earth Space Sci.}{11}{}{e2023EA002965}.
\newblock
\begin{APACrefDOI} \doi{10.1029/2023EA002965} \end{APACrefDOI}
\PrintBackRefs{\CurrentBib}

\bibitem [\protect \citeauthoryear {%
{Bergstedt}%
\ \protect \BOthers {.}}{%
{Bergstedt}%
\ \protect \BOthers {.}}{%
{\protect \APACyear {2020}}%
}]{%
bergstedt20}
\APACinsertmetastar {%
bergstedt20}%
\begin{APACrefauthors}%
{Bergstedt}, K.%
, {Ji}, H.%
, {Jara-Almonte}, J.%
, {Yoo}, J.%
, {Ergun}, R\BPBI E.%
\BCBL {}\ \BBA {} {Chen}, L\BPBI J.%
\end{APACrefauthors}%
\unskip\
\newblock
\APACrefYearMonthDay{2020}{{\APACmonth{10}}}{}.
\newblock
{\BBOQ}\APACrefatitle {{Statistical Properties of Magnetic Structures and
  Energy Dissipation during Turbulent Reconnection in the Earth's Magnetotail}}
  {{Statistical Properties of Magnetic Structures and Energy Dissipation during
  Turbulent Reconnection in the Earth's Magnetotail}}.{\BBCQ}
\newblock
\APACjournalVolNumPages{Geophys. Res. Lett.}{47}{19}{e88540}.
\newblock
\begin{APACrefDOI} \doi{10.1029/2020GL088540} \end{APACrefDOI}
\PrintBackRefs{\CurrentBib}

\bibitem [\protect \citeauthoryear {%
{Bose}%
\ \protect \BOthers {.}}{%
{Bose}%
\ \protect \BOthers {.}}{%
{\protect \APACyear {2024}}%
}]{%
bose2024}
\APACinsertmetastar {%
bose2024}%
\begin{APACrefauthors}%
{Bose}, S.%
, {Fox}, W.%
, {Ji}, H.%
, {Yoo}, J.%
, {Goodman}, A.%
, {Alt}, A.%
\BCBL {}\ \BBA {} {Yamada}, M.%
\end{APACrefauthors}%
\unskip\
\newblock
\APACrefYearMonthDay{2024}{{\APACmonth{05}}}{}.
\newblock
{\BBOQ}\APACrefatitle {{Conversion of Magnetic Energy to Plasma Kinetic Energy
  During Guide Field Magnetic Reconnection in the Laboratory}} {{Conversion of
  Magnetic Energy to Plasma Kinetic Energy During Guide Field Magnetic
  Reconnection in the Laboratory}}.{\BBCQ}
\newblock
\APACjournalVolNumPages{Phys. Rev. Lett.}{132}{20}{205102}.
\newblock
\begin{APACrefDOI} \doi{10.1103/PhysRevLett.132.205102} \end{APACrefDOI}
\PrintBackRefs{\CurrentBib}

\bibitem [\protect \citeauthoryear {%
{Buechner}%
\ \BBA {} {Zelenyi}%
}{%
{Buechner}%
\ \BBA {} {Zelenyi}%
}{%
{\protect \APACyear {1989}}%
}]{%
buechner89}
\APACinsertmetastar {%
buechner89}%
\begin{APACrefauthors}%
{Buechner}, J.%
\BCBT {}\ \BBA {} {Zelenyi}, L\BPBI M.%
\end{APACrefauthors}%
\unskip\
\newblock
\APACrefYearMonthDay{1989}{{\APACmonth{09}}}{}.
\newblock
{\BBOQ}\APACrefatitle {{Regular and chaotic charged particle motion in magnetotaillike field reversals. 1. Basic theory of trapped motion}} {{Regular and chaotic charged particle motion in magnetotaillike field reversals. 1. Basic theory of trapped motion}}.{\BBCQ}
\newblock
\APACjournalVolNumPages{Journal of Geophysical Research: Space Physics}{94}{A9}{11821-11842}.
\newblock
\begin{APACrefDOI} \doi{10.1029/JA094iA09p11821} \end{APACrefDOI}
\PrintBackRefs{\CurrentBib}

\bibitem [\protect \citeauthoryear {%
{Burch}%
\ \protect \BOthers {.}}{%
{Burch}%
\ \protect \BOthers {.}}{%
{\protect \APACyear {2016}}%
}]{%
burch16}
\APACinsertmetastar {%
burch16}%
\begin{APACrefauthors}%
{Burch}, J\BPBI L.%
, {Torbert}, R\BPBI B.%
, {Phan}, T\BPBI D.%
, {Chen}, L\BHBI J.%
, {Moore}, T\BPBI E.%
, {Ergun}, R\BPBI E.%
\BDBL {}{Chandler}, M.%
\end{APACrefauthors}%
\unskip\
\newblock
\APACrefYearMonthDay{2016}{{\APACmonth{06}}}{}.
\newblock
{\BBOQ}\APACrefatitle {{Electron-scale measurements of magnetic reconnection in
  space}} {{Electron-scale measurements of magnetic reconnection in
  space}}.{\BBCQ}
\newblock
\APACjournalVolNumPages{Science}{352}{}{aaf2939}.
\newblock
\begin{APACrefDOI} \doi{10.1126/science.aaf2939} \end{APACrefDOI}
\PrintBackRefs{\CurrentBib}

\bibitem [\protect \citeauthoryear {%
{Cerri}%
, {Arzamasskiy}%
\BCBL {}\ \BBA {} {Kunz}%
}{%
{Cerri}%
\ \protect \BOthers {.}}{%
{\protect \APACyear {2021}}%
}]{%
Cerri2021}
\APACinsertmetastar {%
Cerri2021}%
\begin{APACrefauthors}%
{Cerri}, S\BPBI S.%
, {Arzamasskiy}, L.%
\BCBL {}\ \BBA {} {Kunz}, M\BPBI W.%
\end{APACrefauthors}%
\unskip\
\newblock
\APACrefYearMonthDay{2021}{{\APACmonth{08}}}{}.
\newblock
{\BBOQ}\APACrefatitle {{On Stochastic Heating and Its Phase-space Signatures in
  Low-beta Kinetic Turbulence}} {{On Stochastic Heating and Its Phase-space
  Signatures in Low-beta Kinetic Turbulence}}.{\BBCQ}
\newblock
\APACjournalVolNumPages{Astrophys. J.}{916}{2}{120}.
\newblock
\begin{APACrefDOI} \doi{10.3847/1538-4357/abfbde} \end{APACrefDOI}
\PrintBackRefs{\CurrentBib}

\bibitem [\protect \citeauthoryear {%
{Chen}%
\ \protect \BOthers {.}}{%
{Chen}%
\ \protect \BOthers {.}}{%
{\protect \APACyear {2019}}%
}]{%
chen19}
\APACinsertmetastar {%
chen19}%
\begin{APACrefauthors}%
{Chen}, L\BPBI J.%
, {Wang}, S.%
, {Hesse}, M.%
, {Ergun}, R\BPBI E.%
, {Moore}, T.%
, {Giles}, B.%
\BDBL {}{Lindqvist}, P\BPBI A.%
\end{APACrefauthors}%
\unskip\
\newblock
\APACrefYearMonthDay{2019}{{\APACmonth{06}}}{}.
\newblock
{\BBOQ}\APACrefatitle {{Electron Diffusion Regions in Magnetotail Reconnection
  Under Varying Guide Fields}} {{Electron Diffusion Regions in Magnetotail
  Reconnection Under Varying Guide Fields}}.{\BBCQ}
\newblock
\APACjournalVolNumPages{Geophys. Res. Lett.}{46}{12}{6230-6238}.
\newblock
\begin{APACrefDOI} \doi{10.1029/2019GL082393} \end{APACrefDOI}
\PrintBackRefs{\CurrentBib}

\bibitem [\protect \citeauthoryear {%
{Dahlin}%
, {Drake}%
\BCBL {}\ \BBA {} {Swisdak}%
}{%
{Dahlin}%
\ \protect \BOthers {.}}{%
{\protect \APACyear {2014}}%
}]{%
dahlin14}
\APACinsertmetastar {%
dahlin14}%
\begin{APACrefauthors}%
{Dahlin}, J\BPBI T.%
, {Drake}, J\BPBI F.%
\BCBL {}\ \BBA {} {Swisdak}, M.%
\end{APACrefauthors}%
\unskip\
\newblock
\APACrefYearMonthDay{2014}{{\APACmonth{09}}}{}.
\newblock
{\BBOQ}\APACrefatitle {{The mechanisms of electron heating and acceleration
  during magnetic reconnection}} {{The mechanisms of electron heating and
  acceleration during magnetic reconnection}}.{\BBCQ}
\newblock
\APACjournalVolNumPages{Physics of Plasmas}{21}{9}{092304}.
\newblock
\begin{APACrefDOI} \doi{10.1063/1.4894484} \end{APACrefDOI}
\PrintBackRefs{\CurrentBib}

\bibitem [\protect \citeauthoryear {%
{Dahlin}%
, {Drake}%
\BCBL {}\ \BBA {} {Swisdak}%
}{%
{Dahlin}%
\ \protect \BOthers {.}}{%
{\protect \APACyear {2017}}%
}]{%
dahlin17}
\APACinsertmetastar {%
dahlin17}%
\begin{APACrefauthors}%
{Dahlin}, J\BPBI T.%
, {Drake}, J\BPBI F.%
\BCBL {}\ \BBA {} {Swisdak}, M.%
\end{APACrefauthors}%
\unskip\
\newblock
\APACrefYearMonthDay{2017}{{\APACmonth{09}}}{}.
\newblock
{\BBOQ}\APACrefatitle {{The role of three-dimensional transport in driving
  enhanced electron acceleration during magnetic reconnection}} {{The role of
  three-dimensional transport in driving enhanced electron acceleration during
  magnetic reconnection}}.{\BBCQ}
\newblock
\APACjournalVolNumPages{Physics of Plasmas}{24}{9}{092110}.
\newblock
\begin{APACrefDOI} \doi{10.1063/1.4986211} \end{APACrefDOI}
\PrintBackRefs{\CurrentBib}

\bibitem [\protect \citeauthoryear {%
{Drake}%
, {Swisdak}%
, {Che}%
\BCBL {}\ \BBA {} {Shay}%
}{%
{Drake}%
\ \protect \BOthers {.}}{%
{\protect \APACyear {2006}}%
}]{%
drake06}
\APACinsertmetastar {%
drake06}%
\begin{APACrefauthors}%
{Drake}, J\BPBI F.%
, {Swisdak}, M.%
, {Che}, H.%
\BCBL {}\ \BBA {} {Shay}, M\BPBI A.%
\end{APACrefauthors}%
\unskip\
\newblock
\APACrefYearMonthDay{2006}{{\APACmonth{10}}}{}.
\newblock
{\BBOQ}\APACrefatitle {{Electron acceleration from contracting magnetic islands
  during reconnection}} {{Electron acceleration from contracting magnetic
  islands during reconnection}}.{\BBCQ}
\newblock
\APACjournalVolNumPages{Nature}{443}{}{553-556}.
\newblock
\begin{APACrefDOI} \doi{10.1038/nature05116} \end{APACrefDOI}
\PrintBackRefs{\CurrentBib}

\bibitem [\protect \citeauthoryear {%
{Ergun}%
\ \protect \BOthers {.}}{%
{Ergun}%
\ \protect \BOthers {.}}{%
{\protect \APACyear {2020}}%
}]{%
ergun20}
\APACinsertmetastar {%
ergun20}%
\begin{APACrefauthors}%
{Ergun}, R\BPBI E.%
, {Ahmadi}, N.%
, {Kromyda}, L.%
, {Schwartz}, S\BPBI J.%
, {Chasapis}, A.%
, {Hoilijoki}, S.%
\BDBL {}{Giles}, B\BPBI L.%
\end{APACrefauthors}%
\unskip\
\newblock
\APACrefYearMonthDay{2020}{{\APACmonth{08}}}{}.
\newblock
{\BBOQ}\APACrefatitle {{Observations of Particle Acceleration in Magnetic
  Reconnection-driven Turbulence}} {{Observations of Particle Acceleration in
  Magnetic Reconnection-driven Turbulence}}.{\BBCQ}
\newblock
\APACjournalVolNumPages{Astrophys. J.}{898}{2}{154}.
\newblock
\begin{APACrefDOI} \doi{10.3847/1538-4357/ab9ab6} \end{APACrefDOI}
\PrintBackRefs{\CurrentBib}

\bibitem [\protect \citeauthoryear {%
{Eriksson}%
\ \protect \BOthers {.}}{%
{Eriksson}%
\ \protect \BOthers {.}}{%
{\protect \APACyear {2024}}%
}]{%
Eriksson2024E}
\APACinsertmetastar {%
Eriksson2024E}%
\begin{APACrefauthors}%
{Eriksson}, S.%
, {Ahmadi}, N.%
, {Burch}, J\BPBI L.%
, {Genestreti}, K\BPBI J.%
, {Swisdak}, M.%
, {Argall}, M\BPBI R.%
\BCBL {}\ \BBA {} {Newman}, D\BPBI L.%
\end{APACrefauthors}%
\unskip\
\newblock
\APACrefYearMonthDay{2024}{{\APACmonth{06}}}{}.
\newblock
{\BBOQ}\APACrefatitle {{MMS Observations of Oscillating Energy Conversion and
  Electron Vorticity in an Electron-Scale Layer Within a Southward Magnetopause
  Reconnection Exhaust}} {{MMS Observations of Oscillating Energy Conversion
  and Electron Vorticity in an Electron-Scale Layer Within a Southward
  Magnetopause Reconnection Exhaust}}.{\BBCQ}
\newblock
\APACjournalVolNumPages{Geophys. Res. Lett.}{51}{12}{e2024GL109878}.
\newblock
\begin{APACrefDOI} \doi{10.1029/2024GL109878} \end{APACrefDOI}
\PrintBackRefs{\CurrentBib}

\bibitem [\protect \citeauthoryear {%
{Fermo}%
, {Drake}%
\BCBL {}\ \BBA {} {Swisdak}%
}{%
{Fermo}%
\ \protect \BOthers {.}}{%
{\protect \APACyear {2010}}%
}]{%
fermo10}
\APACinsertmetastar {%
fermo10}%
\begin{APACrefauthors}%
{Fermo}, R\BPBI L.%
, {Drake}, J\BPBI F.%
\BCBL {}\ \BBA {} {Swisdak}, M.%
\end{APACrefauthors}%
\unskip\
\newblock
\APACrefYearMonthDay{2010}{{\APACmonth{01}}}{}.
\newblock
{\BBOQ}\APACrefatitle {{A statistical model of magnetic islands in a current
  layer}} {{A statistical model of magnetic islands in a current
  layer}}.{\BBCQ}
\newblock
\APACjournalVolNumPages{Phys. Plasmas}{17}{1}{010702}.
\PrintBackRefs{\CurrentBib}

\bibitem [\protect \citeauthoryear {%
{Fermo}%
, {Drake}%
, {Swisdak}%
\BCBL {}\ \BBA {} {Hwang}%
}{%
{Fermo}%
\ \protect \BOthers {.}}{%
{\protect \APACyear {2011}}%
}]{%
fermo11}
\APACinsertmetastar {%
fermo11}%
\begin{APACrefauthors}%
{Fermo}, R\BPBI L.%
, {Drake}, J\BPBI F.%
, {Swisdak}, M.%
\BCBL {}\ \BBA {} {Hwang}, K\BPBI J.%
\end{APACrefauthors}%
\unskip\
\newblock
\APACrefYearMonthDay{2011}{{\APACmonth{09}}}{}.
\newblock
{\BBOQ}\APACrefatitle {{Comparison of a statistical model for magnetic islands
  in large current layers with Hall MHD simulations and Cluster FTE
  observations}} {{Comparison of a statistical model for magnetic islands in
  large current layers with Hall MHD simulations and Cluster FTE
  observations}}.{\BBCQ}
\newblock
\APACjournalVolNumPages{Journal of Geophysical Research (Space
  Physics)}{116}{A9}{A09226}.
\newblock
\begin{APACrefDOI} \doi{10.1029/2010JA016271} \end{APACrefDOI}
\PrintBackRefs{\CurrentBib}

\bibitem [\protect \citeauthoryear {%
{Ji}%
\ \protect \BOthers {.}}{%
{Ji}%
\ \protect \BOthers {.}}{%
{\protect \APACyear {2022}}%
}]{%
ji22}
\APACinsertmetastar {%
ji22}%
\begin{APACrefauthors}%
{Ji}, H.%
, {Daughton}, W.%
, {Jara-Almonte}, J.%
, {Le}, A.%
, {Stanier}, A.%
\BCBL {}\ \BBA {} {Yoo}, J.%
\end{APACrefauthors}%
\unskip\
\newblock
\APACrefYearMonthDay{2022}{{\APACmonth{02}}}{}.
\newblock
{\BBOQ}\APACrefatitle {{Magnetic reconnection in the era of exascale computing
  and multiscale experiments}} {{Magnetic reconnection in the era of exascale
  computing and multiscale experiments}}.{\BBCQ}
\newblock
\APACjournalVolNumPages{Nature Reviews Physics}{4}{4}{263-282}.
\newblock
\begin{APACrefDOI} \doi{10.1038/s42254-021-00419-x} \end{APACrefDOI}
\PrintBackRefs{\CurrentBib}

\bibitem [\protect \citeauthoryear {%
{Jiang}%
\ \protect \BOthers {.}}{%
{Jiang}%
\ \protect \BOthers {.}}{%
{\protect \APACyear {2021}}%
}]{%
jiang21}
\APACinsertmetastar {%
jiang21}%
\begin{APACrefauthors}%
{Jiang}, K.%
, {Huang}, S\BPBI Y.%
, {Yuan}, Z\BPBI G.%
, {Deng}, X\BPBI H.%
, {Wei}, Y\BPBI Y.%
, {Xiong}, Q\BPBI Y.%
\BDBL {}{Yu}, L.%
\end{APACrefauthors}%
\unskip\
\newblock
\APACrefYearMonthDay{2021}{{\APACmonth{06}}}{}.
\newblock
{\BBOQ}\APACrefatitle {{Statistical Properties of Current, Energy Conversion,
  and Electron Acceleration in Flux Ropes in the Terrestrial Magnetotail}}
  {{Statistical Properties of Current, Energy Conversion, and Electron
  Acceleration in Flux Ropes in the Terrestrial Magnetotail}}.{\BBCQ}
\newblock
\APACjournalVolNumPages{Geophys. Res. Lett.}{48}{11}{e93458}.
\newblock
\begin{APACrefDOI} \doi{10.1029/2021GL093458} \end{APACrefDOI}
\PrintBackRefs{\CurrentBib}

\bibitem [\protect \citeauthoryear {%
{Lemoine}%
}{%
{Lemoine}%
}{%
{\protect \APACyear {2022}}%
}]{%
lemoine22}
\APACinsertmetastar {%
lemoine22}%
\begin{APACrefauthors}%
{Lemoine}, M.%
\end{APACrefauthors}%
\unskip\
\newblock
\APACrefYearMonthDay{2022}{{\APACmonth{11}}}{}.
\newblock
{\BBOQ}\APACrefatitle {{First-Principles Fermi Acceleration in Magnetized
  Turbulence}} {{First-Principles Fermi Acceleration in Magnetized
  Turbulence}}.{\BBCQ}
\newblock
\APACjournalVolNumPages{Phys. Rev. Lett.}{129}{21}{215101}.
\newblock
\begin{APACrefDOI} \doi{10.1103/PhysRevLett.129.215101} \end{APACrefDOI}
\PrintBackRefs{\CurrentBib}

\bibitem [\protect \citeauthoryear {%
{Li}%
, {Guo}%
, {Li}%
\BCBL {}\ \BBA {} {Li}%
}{%
{Li}%
\ \protect \BOthers {.}}{%
{\protect \APACyear {2015}}%
}]{%
Li2015}
\APACinsertmetastar {%
Li2015}%
\begin{APACrefauthors}%
{Li}, X.%
, {Guo}, F.%
, {Li}, H.%
\BCBL {}\ \BBA {} {Li}, G.%
\end{APACrefauthors}%
\unskip\
\newblock
\APACrefYearMonthDay{2015}{{\APACmonth{10}}}{}.
\newblock
{\BBOQ}\APACrefatitle {{Nonthermally Dominated Electron Acceleration during
  Magnetic Reconnection in a Low-{\ensuremath{\beta}} Plasma}} {{Nonthermally
  Dominated Electron Acceleration during Magnetic Reconnection in a
  Low-{\ensuremath{\beta}} Plasma}}.{\BBCQ}
\newblock
\APACjournalVolNumPages{Astrophys. J. Lett}{811}{2}{L24}.
\newblock
\begin{APACrefDOI} \doi{10.1088/2041-8205/811/2/L24} \end{APACrefDOI}
\PrintBackRefs{\CurrentBib}

\bibitem [\protect \citeauthoryear {%
{Li}%
, {Guo}%
, {Li}%
\BCBL {}\ \BBA {} {Li}%
}{%
{Li}%
\ \protect \BOthers {.}}{%
{\protect \APACyear {2017}}%
}]{%
Li2017}
\APACinsertmetastar {%
Li2017}%
\begin{APACrefauthors}%
{Li}, X.%
, {Guo}, F.%
, {Li}, H.%
\BCBL {}\ \BBA {} {Li}, G.%
\end{APACrefauthors}%
\unskip\
\newblock
\APACrefYearMonthDay{2017}{{\APACmonth{07}}}{}.
\newblock
{\BBOQ}\APACrefatitle {{Particle Acceleration during Magnetic Reconnection in a
  Low-beta Plasma}} {{Particle Acceleration during Magnetic Reconnection in a
  Low-beta Plasma}}.{\BBCQ}
\newblock
\APACjournalVolNumPages{Astrophys. J.}{843}{1}{21}.
\newblock
\begin{APACrefDOI} \doi{10.3847/1538-4357/aa745e} \end{APACrefDOI}
\PrintBackRefs{\CurrentBib}

\bibitem [\protect \citeauthoryear {%
{Ma}%
, {Zhou}%
, {Zhong}%
\BCBL {}\ \BBA {} {Deng}%
}{%
{Ma}%
\ \protect \BOthers {.}}{%
{\protect \APACyear {2022}}%
}]{%
ma22}
\APACinsertmetastar {%
ma22}%
\begin{APACrefauthors}%
{Ma}, W.%
, {Zhou}, M.%
, {Zhong}, Z.%
\BCBL {}\ \BBA {} {Deng}, X.%
\end{APACrefauthors}%
\unskip\
\newblock
\APACrefYearMonthDay{2022}{{\APACmonth{06}}}{}.
\newblock
{\BBOQ}\APACrefatitle {{Contrasting the Mechanisms of Reconnection-driven
  Electron Acceleration with In Situ Observations from MMS in the Terrestrial
  Magnetotail}} {{Contrasting the Mechanisms of Reconnection-driven Electron
  Acceleration with In Situ Observations from MMS in the Terrestrial
  Magnetotail}}.{\BBCQ}
\newblock
\APACjournalVolNumPages{Astrophys. J.}{931}{2}{135}.
\newblock
\begin{APACrefDOI} \doi{10.3847/1538-4357/ac6be6} \end{APACrefDOI}
\PrintBackRefs{\CurrentBib}

\bibitem [\protect \citeauthoryear {%
{Montag}%
, {Egedal}%
, {Lichko}%
\BCBL {}\ \BBA {} {Wetherton}%
}{%
{Montag}%
\ \protect \BOthers {.}}{%
{\protect \APACyear {2017}}%
}]{%
Montag2017}
\APACinsertmetastar {%
Montag2017}%
\begin{APACrefauthors}%
{Montag}, P.%
, {Egedal}, J.%
, {Lichko}, E.%
\BCBL {}\ \BBA {} {Wetherton}, B.%
\end{APACrefauthors}%
\unskip\
\newblock
\APACrefYearMonthDay{2017}{{\APACmonth{06}}}{}.
\newblock
{\BBOQ}\APACrefatitle {{Impact of compressibility and a guide field on Fermi
  acceleration during magnetic island coalescence}} {{Impact of compressibility
  and a guide field on Fermi acceleration during magnetic island
  coalescence}}.{\BBCQ}
\newblock
\APACjournalVolNumPages{Physics of Plasmas}{24}{6}{062906}.
\newblock
\begin{APACrefDOI} \doi{10.1063/1.4985302} \end{APACrefDOI}
\PrintBackRefs{\CurrentBib}

\bibitem [\protect \citeauthoryear {%
Northrop%
}{%
Northrop%
}{%
{\protect \APACyear {1963}}%
}]{%
northrop63}
\APACinsertmetastar {%
northrop63}%
\begin{APACrefauthors}%
Northrop, T.%
\end{APACrefauthors}%
\unskip\
\newblock
\APACrefYearMonthDay{1963}{}{}.
\newblock
{\BBOQ}\APACrefatitle {Adiabatic Charged-Particle Motion} {Adiabatic
  charged-particle motion}.{\BBCQ}
\newblock
\APACjournalVolNumPages{Reviews of Geophysics}{1}{}{283}.
\PrintBackRefs{\CurrentBib}

\bibitem [\protect \citeauthoryear {%
{Pyakurel}%
\ \protect \BOthers {.}}{%
{Pyakurel}%
\ \protect \BOthers {.}}{%
{\protect \APACyear {2025}}%
}]{%
Pyakurel2025}
\APACinsertmetastar {%
Pyakurel2025}%
\begin{APACrefauthors}%
{Pyakurel}, P\BPBI S.%
, {Phan}, T\BPBI D.%
, {{\O}ieroset}, M.%
, {Drake}, J\BPBI F.%
, {Shay}, M\BPBI A.%
, {Mallet}, A.%
\BDBL {}{Strangeway}, R\BPBI J.%
\end{APACrefauthors}%
\unskip\
\newblock
\APACrefYearMonthDay{2025}{{\APACmonth{03}}}{}.
\newblock
{\BBOQ}\APACrefatitle {{Detection of Large Guide Field Electron-Only
  Reconnection in a Filamentary Current Sheet Immersed in a Large-Scale
  Magnetopause Reconnection Exhaust}} {{Detection of Large Guide Field
  Electron-Only Reconnection in a Filamentary Current Sheet Immersed in a
  Large-Scale Magnetopause Reconnection Exhaust}}.{\BBCQ}
\newblock
\APACjournalVolNumPages{Phys. Rev. Lett.}{134}{11}{115201}.
\newblock
\begin{APACrefDOI} \doi{10.1103/PhysRevLett.134.115201} \end{APACrefDOI}
\PrintBackRefs{\CurrentBib}

\bibitem [\protect \citeauthoryear {%
{Shi}%
\ \protect \BOthers {.}}{%
{Shi}%
\ \protect \BOthers {.}}{%
{\protect \APACyear {2006}}%
}]{%
shi06}
\APACinsertmetastar {%
shi06}%
\begin{APACrefauthors}%
{Shi}, Q\BPBI Q.%
, {Shen}, C.%
, {Dunlop}, M\BPBI W.%
, {Pu}, Z\BPBI Y.%
, {Zong}, Q\BPBI G.%
, {Liu}, Z\BPBI X.%
\BDBL {}{Balogh}, A.%
\end{APACrefauthors}%
\unskip\
\newblock
\APACrefYearMonthDay{2006}{{\APACmonth{04}}}{}.
\newblock
{\BBOQ}\APACrefatitle {{Motion of observed structures calculated from
  multi-point magnetic field measurements: Application to Cluster}} {{Motion of
  observed structures calculated from multi-point magnetic field measurements:
  Application to Cluster}}.{\BBCQ}
\newblock
\APACjournalVolNumPages{Geophys. Res. Lett.}{33}{8}{L08109}.
\newblock
\begin{APACrefDOI} \doi{10.1029/2005GL025073} \end{APACrefDOI}
\PrintBackRefs{\CurrentBib}

\bibitem [\protect \citeauthoryear {%
Smith%
, Sun%
, Slavin%
\BCBL {}\ \BBA {} Rae%
}{%
Smith%
\ \protect \BOthers {.}}{%
{\protect \APACyear {2024}}%
}]{%
smith24}
\APACinsertmetastar {%
smith24}%
\begin{APACrefauthors}%
Smith, A\BPBI W.%
, Sun, W.%
, Slavin, J\BPBI A.%
\BCBL {}\ \BBA {} Rae, I\BPBI J.%
\end{APACrefauthors}%
\unskip\
\newblock
\APACrefYearMonthDay{2024}{3}{}.
\newblock
{\BBOQ}\APACrefatitle {Ion-Scale Magnetic Flux Ropes and Loops in Earth's
  Magnetotail: An Automated, Comprehensive Survey of MMS Data Between 2017 and
  2022} {Ion-scale magnetic flux ropes and loops in earth's magnetotail: An
  automated, comprehensive survey of mms data between 2017 and 2022}.{\BBCQ}
\newblock
\APACjournalVolNumPages{Journal of Geophysical Research: Space
  Physics}{129}{}{}.
\newblock
\begin{APACrefDOI} \doi{10.1029/2023JA032231} \end{APACrefDOI}
\PrintBackRefs{\CurrentBib}

\bibitem [\protect \citeauthoryear {%
{Steinvall}%
, {Richard}%
, {F{\"u}l{\"o}p}%
, {Hanebring}%
\BCBL {}\ \BBA {} {Pusztai}%
}{%
{Steinvall}%
\ \protect \BOthers {.}}{%
{\protect \APACyear {2025}}%
}]{%
steinvall2025}
\APACinsertmetastar {%
steinvall2025}%
\begin{APACrefauthors}%
{Steinvall}, K.%
, {Richard}, L.%
, {F{\"u}l{\"o}p}, T.%
, {Hanebring}, L.%
\BCBL {}\ \BBA {} {Pusztai}, I.%
\end{APACrefauthors}%
\unskip\
\newblock
\APACrefYearMonthDay{2025}{{\APACmonth{06}}}{}.
\newblock
{\BBOQ}\APACrefatitle {{Eulerian and Lagrangian electron energisation during
  magnetic reconnection}} {{Eulerian and Lagrangian electron energisation
  during magnetic reconnection}}.{\BBCQ}
\newblock
\APACjournalVolNumPages{Journal of Plasma Physics}{91}{3}{E90}.
\newblock
\begin{APACrefDOI} \doi{10.1017/S0022377825100408} \end{APACrefDOI}
\PrintBackRefs{\CurrentBib}

\bibitem [\protect \citeauthoryear {%
{Sun}%
\ \protect \BOthers {.}}{%
{Sun}%
\ \protect \BOthers {.}}{%
{\protect \APACyear {2022}}%
}]{%
sun22}
\APACinsertmetastar {%
sun22}%
\begin{APACrefauthors}%
{Sun}, W.%
, {Turner}, D\BPBI L.%
, {Zhang}, Q.%
, {Wang}, S.%
, {Egedal}, J.%
, {Leonard}, T.%
\BDBL {}{Burch}, J\BPBI L.%
\end{APACrefauthors}%
\unskip\
\newblock
\APACrefYearMonthDay{2022}{{\APACmonth{12}}}{}.
\newblock
{\BBOQ}\APACrefatitle {{Properties and Acceleration Mechanisms of Electrons Up
  To 200 keV Associated With a Flux Rope Pair and Reconnection X-Lines Around
  It in Earth's Plasma Sheet}} {{Properties and Acceleration Mechanisms of
  Electrons Up To 200 keV Associated With a Flux Rope Pair and Reconnection
  X-Lines Around It in Earth's Plasma Sheet}}.{\BBCQ}
\newblock
\APACjournalVolNumPages{Journal of Geophysical Research (Space
  Physics)}{127}{12}{e2022JA030721}.
\newblock
\begin{APACrefDOI} \doi{10.1029/2022JA030721} \end{APACrefDOI}
\PrintBackRefs{\CurrentBib}

\bibitem [\protect \citeauthoryear {%
{Torbert}%
\ \protect \BOthers {.}}{%
{Torbert}%
\ \protect \BOthers {.}}{%
{\protect \APACyear {2018}}%
}]{%
torbert18}
\APACinsertmetastar {%
torbert18}%
\begin{APACrefauthors}%
{Torbert}, R\BPBI B.%
, {Burch}, J\BPBI L.%
, {Phan}, T\BPBI D.%
, {Hesse}, M.%
, {Argall}, M\BPBI R.%
, {Shuster}, J.%
\BDBL {}{Saito}, Y.%
\end{APACrefauthors}%
\unskip\
\newblock
\APACrefYearMonthDay{2018}{{\APACmonth{12}}}{}.
\newblock
{\BBOQ}\APACrefatitle {{Electron-scale dynamics of the diffusion region during
  symmetric magnetic reconnection in space}} {{Electron-scale dynamics of the
  diffusion region during symmetric magnetic reconnection in space}}.{\BBCQ}
\newblock
\APACjournalVolNumPages{Science}{362}{6421}{1391-1395}.
\newblock
\begin{APACrefDOI} \doi{10.1126/science.aat2998} \end{APACrefDOI}
\PrintBackRefs{\CurrentBib}

\bibitem [\protect \citeauthoryear {%
{Wan}%
\ \protect \BOthers {.}}{%
{Wan}%
\ \protect \BOthers {.}}{%
{\protect \APACyear {2015}}%
}]{%
Wan2015}
\APACinsertmetastar {%
Wan2015}%
\begin{APACrefauthors}%
{Wan}, M.%
, {Matthaeus}, W\BPBI H.%
, {Roytershteyn}, V.%
, {Karimabadi}, H.%
, {Parashar}, T.%
, {Wu}, P.%
\BCBL {}\ \BBA {} {Shay}, M.%
\end{APACrefauthors}%
\unskip\
\newblock
\APACrefYearMonthDay{2015}{{\APACmonth{05}}}{}.
\newblock
{\BBOQ}\APACrefatitle {{Intermittent Dissipation and Heating in 3D Kinetic
  Plasma Turbulence}} {{Intermittent Dissipation and Heating in 3D Kinetic
  Plasma Turbulence}}.{\BBCQ}
\newblock
\APACjournalVolNumPages{Phys. Rev. Lett.}{114}{17}{175002}.
\newblock
\begin{APACrefDOI} \doi{10.1103/PhysRevLett.114.175002} \end{APACrefDOI}
\PrintBackRefs{\CurrentBib}

\bibitem [\protect \citeauthoryear {%
{Wilder}%
\ \protect \BOthers {.}}{%
{Wilder}%
\ \protect \BOthers {.}}{%
{\protect \APACyear {2018}}%
}]{%
wilder18}
\APACinsertmetastar {%
wilder18}%
\begin{APACrefauthors}%
{Wilder}, F\BPBI D.%
, {Ergun}, R\BPBI E.%
, {Burch}, J\BPBI L.%
, {Ahmadi}, N.%
, {Eriksson}, S.%
, {Phan}, T\BPBI D.%
\BDBL {}{Khotyaintsev}, Y\BPBI V.%
\end{APACrefauthors}%
\unskip\
\newblock
\APACrefYearMonthDay{2018}{{\APACmonth{08}}}{}.
\newblock
{\BBOQ}\APACrefatitle {{The Role of the Parallel Electric Field in
  Electron-Scale Dissipation at Reconnecting Currents in the Magnetosheath}}
  {{The Role of the Parallel Electric Field in Electron-Scale Dissipation at
  Reconnecting Currents in the Magnetosheath}}.{\BBCQ}
\newblock
\APACjournalVolNumPages{Journal of Geophysical Research (Space
  Physics)}{123}{8}{6533-6547}.
\newblock
\begin{APACrefDOI} \doi{10.1029/2018JA025529} \end{APACrefDOI}
\PrintBackRefs{\CurrentBib}

\bibitem [\protect \citeauthoryear {%
{Xu}%
\ \protect \BOthers {.}}{%
{Xu}%
\ \protect \BOthers {.}}{%
{\protect \APACyear {2023}}%
}]{%
xu23}
\APACinsertmetastar {%
xu23}%
\begin{APACrefauthors}%
{Xu}, Q.%
, {Zhou}, M.%
, {Ma}, W.%
, {He}, J.%
, {Huang}, S.%
, {Zhong}, Z.%
\BDBL {}{Deng}, X.%
\end{APACrefauthors}%
\unskip\
\newblock
\APACrefYearMonthDay{2023}{{\APACmonth{03}}}{}.
\newblock
{\BBOQ}\APACrefatitle {{Electron Heating in Magnetosheath Turbulence: Dominant
  Role of the Parallel Electric Field Within Coherent Structures}} {{Electron
  Heating in Magnetosheath Turbulence: Dominant Role of the Parallel Electric
  Field Within Coherent Structures}}.{\BBCQ}
\newblock
\APACjournalVolNumPages{Geophys. Res. Lett.}{50}{6}{e2022GL102523}.
\newblock
\begin{APACrefDOI} \doi{10.1029/2022GL102523} \end{APACrefDOI}
\PrintBackRefs{\CurrentBib}

\bibitem [\protect \citeauthoryear {%
{Yamada}%
, {Kulsrud}%
\BCBL {}\ \BBA {} {Ji}%
}{%
{Yamada}%
\ \protect \BOthers {.}}{%
{\protect \APACyear {2010}}%
}]{%
yamada10}
\APACinsertmetastar {%
yamada10}%
\begin{APACrefauthors}%
{Yamada}, M.%
, {Kulsrud}, R.%
\BCBL {}\ \BBA {} {Ji}, H.%
\end{APACrefauthors}%
\unskip\
\newblock
\APACrefYearMonthDay{2010}{{\APACmonth{01}}}{}.
\newblock
{\BBOQ}\APACrefatitle {{Magnetic reconnection}} {{Magnetic
  reconnection}}.{\BBCQ}
\newblock
\APACjournalVolNumPages{Reviews of Modern Physics}{82}{1}{603-664}.
\newblock
\begin{APACrefDOI} \doi{10.1103/RevModPhys.82.603} \end{APACrefDOI}
\PrintBackRefs{\CurrentBib}

\bibitem [\protect \citeauthoryear {%
{Zenitani}%
, {Hesse}%
, {Klimas}%
, {Black}%
\BCBL {}\ \BBA {} {Kuznetsova}%
}{%
{Zenitani}%
\ \protect \BOthers {.}}{%
{\protect \APACyear {2011}}%
}]{%
Zenitani2011}
\APACinsertmetastar {%
Zenitani2011}%
\begin{APACrefauthors}%
{Zenitani}, S.%
, {Hesse}, M.%
, {Klimas}, A.%
, {Black}, C.%
\BCBL {}\ \BBA {} {Kuznetsova}, M.%
\end{APACrefauthors}%
\unskip\
\newblock
\APACrefYearMonthDay{2011}{{\APACmonth{12}}}{}.
\newblock
{\BBOQ}\APACrefatitle {{The inner structure of collisionless magnetic
  reconnection: The electron-frame dissipation measure and Hall fields}} {{The
  inner structure of collisionless magnetic reconnection: The electron-frame
  dissipation measure and Hall fields}}.{\BBCQ}
\newblock
\APACjournalVolNumPages{Physics of Plasmas}{18}{12}{122108-122108}.
\newblock
\begin{APACrefDOI} \doi{10.1063/1.3662430} \end{APACrefDOI}
\PrintBackRefs{\CurrentBib}

\end{thebibliography}
\end{document}